\documentclass[11pt]{article}
\usepackage{amsmath,amssymb,amsfonts}
\usepackage[T2A]{fontenc}

\newtheorem{theorem}{Theorem}
\newtheorem{proposition}{Proposition}

\def\D{{\cal D}}

\def\R{{\mathbb R}}
\def\C{{\mathbb C}}

\def\Q{{\mathbb H}}
\def\n{{\bf n}}
\def\x{{\bf x}}
\def\X{{\bf X}}
\def\H{{\bf H}}

\def\p{{\bf p}}
\def\Re{{\mathrm{Re}\, }}
\def\Im{{\mathrm{Im}\, }}

\begin{document}

\title{The Moutard transformation of two-dimensional Dirac operators and
conformal geometry of surfaces in the four-space
\thanks{The work was supported by the grant 3485/GF4 of Ministry of education and science of Republic of Kazakhstan.}}
\author{R.M. Matuev
\thanks{Sobolev Institute of Mathematics, Academician Koptyug avenue 4, 630090, Novosibirsk, Russia, and
Department of Mathematics and Mechanics, Novosibirsk State University, Pirogov street 2, 630090 Novosibirsk, Russia;
e-mail: rmatuev@yandex.ru.}
\and I.A. Taimanov
\thanks{Sobolev Institute of Mathematics, Academician Koptyug avenue 4, 630090, Novosibirsk, Russia, and
Department of Mathematics and Mechanics, Novosibirsk State University, Pirogov street 2, 630090 Novosibirsk, Russia;
e-mail: taimanov@math.nsc.ru. }}
\date{}
\maketitle

\section{Introduction}

In \cite{T151} the Moutard transformation, for a two-dimensional Dirac operator with a real-valued potential,
derived in \cite{C}, was related with conformal geometry of surfaces in the three-space.
In this article we expand this picture for surfaces in the four-space, because every such a surface
admits a Weierstrass representation related to a two-dimensional Dirac operator \cite{TDS,T-RS}.

Therewith we generalize the Moutard transformation from \cite{C} onto Dirac operators with complex-valued potentials, i.e.
for operators of the form
\begin{equation}
\label{dirac}
\cal D = \left(
\begin{array}{cc}
0 & \partial \\
-\bar{\partial} & 0
\end{array}
\right) + \left(
\begin{array}{cc}
U & 0 \\
0 & \bar{U}
\end{array}
\right),
\end{equation}
where $\partial = \frac{1}{2}\big(\frac{\partial}{\partial x} - i\frac{\partial}{\partial y}\big)$ and
$\bar{\partial} =
\frac{1}{2}\big(\frac{\partial}{\partial x} + i\frac{\partial}{\partial y}\big)$.

Let us briefly expose the main results. We consider the operator
$$
\D^\vee = \left(\begin{array}{cc} 0 & \partial \\
-\bar{\partial} & 0 \end{array}\right) +
\left(\begin{array}{cc} \bar{U} & 0 \\
0 & U \end{array}\right).
$$
Let $\psi$ and $\varphi$ satisfy the equations
\begin{equation}
\label{diraceq}
\D\psi=0, \ \ \ \ \D^\vee\varphi=0.
\end{equation}
Then the matrix-valued functions
\begin{equation}
\label{quaternionsol}
\Psi = \left(\begin{array}{cc} \psi_1 & -\bar{\psi}_2 \\
\psi_2 & \bar{\psi}_1 \end{array} \right), \ \ \ \
\Phi = \left(\begin{array}{cc} \varphi_1 & -\bar{\varphi}_2 \\
\varphi_2 & \bar{\varphi}_1 \end{array} \right)
\end{equation}
satisfy the equations
\begin{equation}
\label{matrixdirac}
{\cal D} \Psi = 0, \ \ \ \D^\vee\Phi = 0,
\end{equation}
which, in fact, means the solutions of (\ref{diraceq}) are invariant with respect to
the transformations$$
\left(\begin{array}{c} \psi_1 \\ \psi_2 \end{array}\right) \longrightarrow
\left(\begin{array}{c} -\bar{\psi}_2 \\ \bar{\psi}_1 \end{array}\right), \ \ \
\left(\begin{array}{c} \varphi_1 \\ \varphi_2 \end{array}\right) \longrightarrow
\left(\begin{array}{c} -\bar{\varphi}_2 \\ \bar{\varphi}_1 \end{array}\right).
$$
Here it is important that the potentials are complex conjugate to each other.

In this article we show that

\begin{enumerate}
\item
{\sl every pair of solutions $\psi$ and $\varphi$ to (\ref{diraceq}) and every point $x_0 \in \R^4$
define a transformation of the Moutard type of the operator $\D$ to an operator of the same form;}

\item
{\sl geometrically the Moutard transformation is given by an action of  composition of the inversion and the reflection, with respect to a line, on a surface in
$\R^4$. This surface is defined via the Weierstrass represen\-tation by vector functions (spinors)
$\psi$ and $\varphi$ and $x_0\in \R^4$ and the potential $U$ of the Dirac operator enters into this Weierstrass represen\-tation.}
\end{enumerate}

\section{The Moutard transformation}
%\label{section3}

Let us consider the quaternion algebra $\Q$, realized by matrices of the form
$\left(\begin{array}{cc} a & b \\ -\bar{b} & \bar{a} \end{array}\right), a,b \in \C$.
To every vector function  $\left(\begin{array}{c} \psi_1 \\ \psi_2 \end{array}\right)$ we correspond a matrix valued function
$$
\Psi = \left(\begin{array}{cc} \psi_1 & -\bar{\psi}_2 \\ \psi_2 & \bar{\psi}_1 \end{array}\right)
$$
with the value in $\Q$.

To every pair $\Phi$ and $\Psi$
of $\Q$-valued functions we correspond the $1$-form $\omega$:
\begin{equation}
\label{omega}
\omega(\Phi,\Psi) = \Phi^\top \Psi dy - i \Phi^\top \sigma_3 \Psi dx
=
\end{equation}
$$
-\frac{i}{2}\left(\Phi^\top \sigma_3 \Psi + \Phi^\top \Psi\right) dz - \frac{i}{2}\left(\Phi^\top \sigma_3 \Psi - \Phi^\top \Psi\right) d\bar{z},
$$
and the function
\begin{equation}
\label{sigma}
S(\Phi,\Psi)(z,\bar{z},t) = \Gamma \int_0^z \omega(\Phi,\Psi),
\end{equation}
where
$$
\Gamma = \left(\begin{array}{cc} 0 & 1 \\ -1 & 0 \end{array}\right) = i\sigma_2, \ \ \
\sigma_3= \left(\begin{array}{cc} 1 & 0 \\ 0 & -1 \end{array}\right),
$$
$\sigma_2$ and $\sigma_3$ are the Pauli matrices.

The form $\omega$ and the function $S$ take values in $\Q$, and moreover $S$ is defined up to integration constants, i.e. up to a constant matrix from
$\Q$.

Here and in the sequel we define the transposition of $X$ by $X^\top$.

To every pair of $\Q$-valued functions $\Phi$ and $\Psi$ we correspond a matrix valued function
\begin{equation}
\label{kmatrix}
K(\Phi,\Psi) =  \Psi S^{-1}(\Phi,\Psi)\Gamma \Phi^\top\Gamma^{-1} =
\left(\begin{array}{cc} i\bar{W} & a \\ -\bar{a} & -iW \end{array}\right).
\end{equation}

By straightforward computations it is proved that

\begin{theorem}
\label{th1}
Let $\Psi_0$  and $\Phi_0$ be solutions of the form (\ref{quaternionsol}) of the Dirac equations
(\ref{diraceq}).

Then for every pair $\Psi$ and $\Phi$  of solutions of (\ref{diraceq})
the functions
\begin{equation}
\label{moutard1}
\begin{split}
\widetilde{\Psi} =  \Psi - \Psi_0 S^{-1}(\Phi_0,\Psi_0) S(\Phi_0,\Psi), \\
\widetilde{\Phi} =  \Phi - \Phi_0 S^{-1}(\Psi_0,\Phi_0) S(\Psi_0,\Phi)
\end{split}
\end{equation}
satisfy the Dirac equations
$$
\widetilde{\D}\widetilde{\Psi} = 0, \ \ \ \ \widetilde{\D}^\vee \widetilde{\Phi} = 0
$$
for the Dirac operators $\widetilde{\D}$ and $\widetilde{\D}^\vee$ with the potential
\begin{equation}
\label{newpotential}
\widetilde{U} = U + W,
\end{equation}
where $W$ is defined by the formula (\ref{kmatrix}) for $K(\Phi_0,\Psi_0)$.
\end{theorem}

{\sc Remarks.} 1)
Due to matrix integration constants in (\ref{sigma}) $\widetilde{\Psi}$ and $\widetilde{\Phi}$ are defined up to multiplication on
$(\Psi_0 S^{-1}(\Phi_0, \Psi_0)) \cdot A$ and $(\Phi_0 S^{-1}(\Psi_0, \Phi_0)) \cdot B$, respectively, with $A$ and $ B$
constant matrices from $\Q$.

2) The formulas (\ref{omega}) are (\ref{sigma}) the same as for the Moutard transformation of the Dirac operator with
a real-valued potential $U$ \cite{T151,C}. The  transformation from Theorem 1 reduces to it for a real-valued potential $U$ and $\Phi_0=\Psi_0$.
The proof of Theorem 1 will follow to its analogue, for the case of a real-valued potential, given in
\cite{T151}.

{\sc Proof.}
1) Let
$$
\widetilde{\Psi}_0 = \Psi_0 S^{-1}(\Phi_0, \Psi_0), \ \ \
\widetilde{\Phi}_0 = \Phi_0 S^{-1}(\Psi_0, \Phi_0).
$$
WE show that $\widetilde{\Psi}_0$ and $\widetilde{\Phi}_0$ satisfy the Dirac equations
\begin{equation}
\label{newdirac}
\widetilde{\D} \widetilde{\Psi} = \D_0 \widetilde{\Psi} + \left(\begin{array}{cc} \widetilde{U} & 0 \\ 0 & \bar{\widetilde{U}}
 \end{array}\right)\widetilde{\Psi} = 0, \ \ \ \widetilde{\D}^\vee\widetilde{\Phi} =0,
\end{equation}
where
$$
\D_0 = \left(\begin{array}{cc} 0 & \partial \\ -\bar{\partial} & 0 \end{array}\right),
$$
with the potential $\widetilde{U} = U+W$ given by (\ref{newpotential}).

Let us apply the ``Leibniz rule'' \cite{T151}
\begin{equation}
\label{leibniz}
\D_0 (A \cdot B) = (\D_0\,A) \cdot B +
\left(\begin{array} {cc} 0 & 1 \\ 0 & 0 \end{array}\right) A \cdot  \partial B
+
\left(\begin{array} {cc} 0 & 0 \\ -1 & 0 \end{array}\right) A \cdot \bar{\partial} B
\end{equation}
to $A = \Psi_0$ and $B= S^{-1} = S^{-1}(\Phi_0, \Psi_0)$:
\begin{equation}
\label{new}
\begin{split}
\D_0 (\Psi_0 S^{-1}(\Phi_0, \Psi_0)) =
\\
(\D_0 \Psi_0)S^{-1} +
\left(\begin{array} {cc} 0 & 1 \\ 0 & 0 \end{array}\right) \Psi_0 S^{-1}_z
+ \left(\begin{array} {cc} 0 & 0 \\ -1 & 0 \end{array}\right) \Psi_0 S^{-1}_{\bar z} =
\\
= - \left(\begin{array}{cc} U & 0 \\ 0 & \bar{U}
 \end{array}\right) \Psi_0 S^{-1}  +
 i \left(\begin{array} {cc} 0 & 1 \\ 0 & 0 \end{array}\right) \Psi_0 S^{-1} \Gamma \Phi_0^\top
 \left(\begin{array}{cc} 1 & 0 \\ 0 & 0 \end{array}\right)\Psi_0  S^{-1} +
\\
 + i \left(\begin{array} {cc} 0 & 0 \\ -1 & 0 \end{array}\right)  \Psi_0 S^{-1}\Gamma \Phi_0^\top
\left(\begin{array} {cc} 0 & 0 \\ 0 & -1 \end{array}\right) \Psi_0 S^{-1}.
\end{split}
\end{equation}
It follows from $S^{-1}S = 1$ that
$$
(S^{-1})_z = - S^{-1} S_z S^{-1},   (S^{-1})_{\bar{z}} = - S^{-1} S_{\bar{z}} S^{-1}
$$
and, by the definition of $S(\Phi_0,\Psi_0)$, we have
\begin{equation}
\label{derivative}
(S^{-1})_z= \frac{i}{2}S^{-1} \Gamma \Phi_0^\top (\sigma_3+1)\Psi_0 S^{-1}, \ \
(S^{-1})_{\bar{z}}= \frac{i}{2} S^{-1} \Gamma \Phi_0^\top (\sigma_3-1)\Psi_0 S^{-1}.
\end{equation}
In view of these identities the formula (\ref{new}) takes the form
$$
\D_0 (\Psi_0 S^{-1}) =
 - \left(\begin{array}{cc} U & 0 \\ 0 & \bar{U}
 \end{array}\right)(\Psi_0 S^{-1}) +
$$
$$
+ i \left(\left(\begin{array} {cc} 0 & 1 \\ 0 & 0 \end{array}\right) G
 \left(\begin{array}{cc} 1 & 0 \\ 0 & 0 \end{array}\right)
+ \left(\begin{array} {cc} 0 & 0 \\ 1 & 0 \end{array}\right) G
\left(\begin{array} {cc} 0 & 0 \\ 0 & 1 \end{array}\right)\right)(\Psi_0 S^{-1}) =
$$
$$
=
 -\left(\begin{array}{cc} U + W& 0 \\ 0 & \bar{U} + \bar{W}
 \end{array}\right)(\Psi_0 S^{-1}),
$$
where
$$
G = K(\Phi_0, \Psi_0) \Gamma^{-1} = \Psi_0 S^{-1}(\Phi_0, \Psi_0) \Gamma \Phi_0^\top = \left(\begin{array} {cc} -a & i\bar{W} \\ iW & -\bar{a} \end{array}\right).
$$
therefore we prove that $\tilde{\Psi}_0$ satisfies the first equation from (\ref{newdirac}).
Analogously it is proved that $\tilde{\Phi}_0$ satisfies the second equation from (\ref{newdirac}).

2)  Let us find a transformation of an arbitrary solution $\Psi$ of (\ref{matrixdirac}) to a solution
$\widetilde{\Psi}$ of (\ref{newdirac}).
WE will look for it in the form
$$
\widetilde{\Psi} = \Psi + \widetilde{\Psi}_0 N.
$$
By (\ref{leibniz}), we have
$$
0 = \widetilde{\D}\widetilde{\Psi} = (\D +\left(\begin{array}{cc} W & 0 \\ 0 & \bar{W} \end{array}\right))(\Psi + \widetilde{\Psi}_0 N) = \D\Psi + \left(\begin{array}{cc} W & 0 \\ 0 & \bar{W} \end{array}\right) \Psi +
(\widetilde{\D}\widetilde{\Psi}_0)\cdot N +
$$
$$
+ \left(\begin{array}{cc} 0 & 1 \\ 0 & 0 \end{array}\right)\widetilde{\Psi}_0 \partial N +
\left(\begin{array}{cc} 0 & 0 \\ -1 & 0 \end{array}\right)\widetilde{\Psi}_0 \bar{\partial} N,
$$
where $W = \widetilde{U}-U$,
and, since  $\widetilde{\D}\widetilde{\Psi} = \D \Psi=0$, we will look for $N$ such that
$$
\left(\begin{array}{cc} W & 0 \\ 0 & \bar{W} \end{array}\right) \Psi = -\left(\begin{array}{cc} 0 & 1 \\ 0 & 0 \end{array}\right)\widetilde{\Psi}_0 \partial N -
\left(\begin{array}{cc} 0 & 0 \\ -1 & 0 \end{array}\right)\widetilde{\Psi}_0 \bar{\partial} N.
$$
However it follows from the formula for $W$ that
$$
\left(\begin{array}{cc} W & 0 \\ 0 & \bar{W} \end{array}\right) \Psi = \left(\begin{array}{cc} 0 & 1 \\ 0 & 0 \end{array}\right)\widetilde{\Psi}_0 S_z (\Phi_0,\Psi) +
\left(\begin{array}{cc} 0 & 0 \\ -1 & 0 \end{array}\right)\widetilde{\Psi}_0 S_{\bar{z}}(\Phi_0,\Psi),
$$
therefore $N$ is equal to
$$
N = -S(\Phi_0,\Psi)
$$
up to a constant matrix from $\Q$
and, hence, the action of the Moutard transformation on $\Psi$ takes the form pointed out by Theorem 1:
$$
\widetilde{\Psi} = \Psi - \Psi_0 S^{-1}(\Phi_0,\Psi_0) S(\Phi_0,\Psi).
$$
Analogously the transformation of $\Phi$ is derived.

Theorem 1 is proved.

\section{Geometry of the Moutard transformation}
\label{section4}

\subsection{The Weierstrass representation of surfaces in $\R^4$}

 The Weierstrass representation of surfaces in $\R^4$ correspond to solutions $\psi$ and $\varphi$ of
(\ref{diraceq}) the surface defined by the formulas
\begin{equation}
\label{int4} x^k(P) = x^k(P_0) + \int \left( x^k_z dz + \bar{x}^k_z
d\bar{z}\right), \ \ k=1,2,3,4,
\end{equation}
where
\begin{equation}
\label{int40}
\begin{split}
x^1_z = \frac{i}{2} (\bar{\varphi}_2\bar{\psi}_2 + \varphi_1
\psi_1), \ \ \ \ x^2_z = \frac{1}{2} (\bar{\varphi}_2\bar{\psi}_2 - \varphi_1 \psi_1),
\\
x^3_z = \frac{1}{2} (\bar{\varphi}_2 \psi_1 + \varphi_1
\bar{\psi}_2), \ \ \ \ x^4_z = \frac{i}{2} (\bar{\varphi}_2 \psi_1 -
\varphi_1 \bar{\psi}_2),
\end{split}
\end{equation}
where the integral is taken along a path from the initial point $P_0$ to $P$.
Therewith the induced metric is equal to
$$
e^{2\alpha} dzd\bar{z} =
(|\psi_1|^2+|\psi_2|^2)(|\varphi_1|^2+|\varphi_2|^2)dz d\bar{z}
$$
and the mean curvature vector
$$
\H = \frac{2 x_{z\bar{z}}}{e^{2\alpha}}
$$
is related to $U$ as follows
$$
|U| = \frac{|\H| e^\alpha}{2}.
$$
These formulas for constructing surfaces in $\R^4$ were introduced in \cite{K2}.
For $\psi=\varphi, U=\bar{U}$ we have $x^4=\mathrm{const}$ and these formulas reduce to analogous
formulas for surfaces in $\R^3$.

These formulas have a local character and for their globalization it is necessary
to consider vector functions $\psi$ as sections of spinor bundles. Such a representation is constructed (up to a multiplication of $\psi$ by $\pm 1$) for every surface 
in $\R^3$ and therewith $4 \int U^2 dx\,dy$ coincides with the value of the Willmore functional \cite{T-MNV}.

For surfaces in $\R^4$ the situation is more complicated \cite{TDS}: every surface in $\R^4$ is also given by such formulas,
however $\psi$ are $\varphi$ are defined by a factorization of the Gauss map and are defined not uniquely but up to gauge transformations
\begin{equation}
\label{gauge}
\left(
\begin{array}{c}
\psi_1 \\ \psi_2
\end{array}
\right) \to \left(
\begin{array}{c}
e^h\psi_1 \\ e^{\bar{h}}\psi_2
\end{array}
\right), \ \ \ \left(
\begin{array}{c}
\varphi_1 \\ \varphi_2
\end{array}
\right) \to \left(
\begin{array}{c}
e^{-h} \varphi_1 \\ e^{-\bar{h}}\varphi_2
\end{array}
\right),
\end{equation}
where $h$ is an arbitrary function on the universal covering of the surface.
Among $\psi$ and $\varphi$, constructed from the Gauss map,
we can find such functions which satisfy the Dirac equations (\ref{diraceq}).  The spinors $\psi$ and $\varphi$,
satisfying (\ref{diraceq}), again are not uniquely defined but up to gauge transformations
(\ref{gauge}) where $h$ is a holomorphic function on the universal covering.
Therewith the phase of the potential is also changed:
\begin{equation}
\label{gaugep}
U \to e^{(\bar{h}-h)}U.
\end{equation}

Let us clarify the relation of $U$ to the mean curvature vector.
For a surface $M$ â $\R^4$ at every point $x \in M$ there exists a two-dimensional space $\nu M_x$ formed by all  tangent vectors to $\R^4$ which are normal to the surface.
Given a Weierstrass representation of the surface, let us choose in
$\nu M_x$ the basis $\n_1$ and $\n_2$ in the form
$$
\n_1 = e^{-\alpha}(-\Im (\psi_2\varphi_1-\bar{\psi}_1\bar{\varphi}_2), -\Re(\bar{\psi}_1\bar{\varphi}_2+\psi_2\varphi_1),
$$
$$
\Re(\psi_2\bar{\varphi}_2-\bar{\psi}_1\varphi_1),-\Im (\bar{\psi}_1\varphi_1+\psi_2\bar{\varphi}_2)),
$$
$$
\n_2 = e^{-\alpha}(\Re(\psi_2\varphi_1-\bar{\psi}_1\bar{\varphi}_2), -\Im(\bar{\psi}_1\bar{\varphi}_2+\psi_2\varphi_1),
$$
$$
\Im (\psi_2\bar{\varphi}_2 - \bar{\psi}_1\varphi_1), \Re (\bar{\psi}_1\varphi_1 + \psi_2\bar{\varphi}_2)).
$$
Define a complex-valued vector
\begin{equation}
\label{normal}
\p = e^{\alpha}(\n_1+i\n_2).
\end{equation}
By straightforward computations, it is shown the the potential $U$ of the Weierstrass representation
takes the form
$$
U = \frac{1}{2}\langle \H,\p\rangle,
$$
where $\langle \cdot,\cdot \rangle$ is the Euclidean (not Hermitian) scalar product.
Different Weier\-strass representations of the same surface with a fixed conformal parameter are related by a gauge transformation
(\ref{gauge}) and, therefore, the vectors $\n_1$ and $\n_2$, constructed from the representations,
as well as $U$ are related by  (\ref{gaugep}).

\subsection{The inversion of $\R^4$ and the Moutard transformation}

By the Liouville theorem, for $n \geq 3$ the group formed by all orientation-preserving
conformal transformations of $S^n = \R^n \cup \{\infty\}$ is generated by translations, rotations of
$\R^n$, and the inversion.

The inversion of $\R^4$ has the form
$$
T: \x \to \frac{\x}{|\x|^2}, \ \ \ \x = (x^1,\dots,x^4) \in \R^4.
$$
In \cite{T151} in the definition of the inversion the right-hand side of the analogous formula was taken
with the opposite sign to preserve the orientation.

Let $u$ be a vector tangent to $\R^4$ at $\x$: $u \in T_{\x}\R^4$ and let $\x\neq 0$.
By straightforward computations we derive the formula
$$
T^\ast u =  \frac{u}{|\x|^2} - 2\x \frac{\langle \x,u\rangle}{|\x|^4}.
$$
This implies that
$$
\langle T^\ast u, T^\ast v\rangle = \frac{\langle u,v \rangle}{|\x|^4}, \ \ \ \ u,v \in T_{\x}\R^4.
$$
Let us consider an immersed surface
$r: {\cal U} \to \R^4$
with a conformal parameter $z$.
The inversion maps it into the surface
$\widetilde{r} = T \cdot r: {\cal U} \to \R^4$,
on which $z$ is also a conformal parameter and the conformal factors of the metrics satisfy the equality
$$
e^{\widetilde{\alpha}(z,\bar{z})} = \frac{e^{\alpha(z,\bar{z})}}{|r(z,\bar{z})|^2}, \ \ \ e^{2\widetilde{\alpha}} =
\frac{1}{2} \langle \widetilde{r}_z, \widetilde{r}_{\bar{z}}\rangle, \ \ e^{2\alpha} = \frac{1}{2} \langle r_z, r_{\bar{z}}\rangle.
$$

Let $\psi = \left(\begin{array}{c} \psi_1 \\ \psi_2 \end{array}\right)$  and $\varphi = \left(\begin{array}{c} \varphi_1 \\ \varphi_2 \end{array}\right)$ define the surface
$r:  {\cal U} \to \R^4$
via the Weierstrass representation.

Let us identify $\R^4$ with the Lie algebra $u(2)$ (or, which that same, with the matrix realization of quaternions)
by the mapping
$$
\x = (x^1,x^2,x^3,x^4) \to
\X = \left(\begin{array}{cc} ix^3 + x^4& -x^1 - ix^2 \\ x^1-ix^2 & -ix^3 +x^4\end{array}\right) .
$$
By straightforward computation,we derive

\begin{proposition}
\label{comp}
In this representation the map
\begin{equation}
\label{stransform}
\X \longrightarrow \X^{-1}
\end{equation}
is a composition of the inversion $\x \to \frac{\x}{|\x|^2}$ and the reflection
\begin{equation}
\label{reflection}
(x^1,x^2,x^3,x^4) \to (-x^1,-x^2,-x^3,x^4).
\end{equation}
\end{proposition}

We have

\begin{proposition}
The formula (\ref{sigma}) gives an immersion into $u(2) = \R^4$ of the surface defined by the spinors
$\psi$ and $\varphi$ via the Weierstrass representation.
\end{proposition}

{\sc Proof.}
By (\ref{int40}) and (\ref{omega}),
$$
S(\Phi_0,\Psi_0)(P) = \Gamma \int_{P_0}^P
-\frac{i}{2}\left(\Phi_0^\top (\sigma_3+1) \Psi_0 dz + \Phi_0^\top (\sigma_3 -1)
\Psi_0\right) d\bar{z}) =
$$
\begin{equation}
\label{surface}
= i \int_0^P \left(\begin{array}{cc} \psi_1 \bar{\varphi}_2 & -\bar{\psi}_2 \bar{\varphi}_2 \\
\psi_1 \varphi_1 & -\bar{\psi}_2 \varphi_1 \end{array}\right)dz +
\left(\begin{array}{cc} \psi_2 \bar{\varphi}_1 & \bar{\psi}_1 \bar{\varphi}_1 \\
-\psi_2 \varphi_2 & -\bar{\psi}_1 \varphi_2 \end{array}\right)d\bar{z} =
\end{equation}
$$
= \int_0^P d \left(\begin{array}{cc} ix^3 + x^4 & -x^1 - ix^2 \\ x^1-ix^2 & -ix^3 + x^4\end{array}\right) \in u(2) ,
$$
i.e. $S$ is the surface determined by the spinors $\psi$ and $\varphi$ via the Weierstrass representation.
Proposition is proved.

The following theorem demonstrates the geometrical meaning of the Moutard transformation from Theorem 1.

\begin{theorem}
Let a surface
$$
S = S(\Phi_0,\Psi_0): {\cal U}  \to \R^4
$$
with a conformal parameter $z \in {\cal U}  \subset \C$ is defined by the spinors
$\Psi_0$ and $\Phi_0$ via the Weierstrass representation.
Then the surface
$$
S^{-1}: {\cal U} \to \R^4 \cup \{\infty\},
$$
obtained from $S$ by applying the composition of the inversion and the reflection (see Proposition \ref{comp})
is defined by the spinors
$$
\widetilde{\Psi}_0 = \Psi_0 S^{-1}(\Phi_0, \Psi_0), \ \ \ \
\widetilde{\Phi}_0 = \Phi_0 S^{-1}(\Psi_0, \Phi_0)
$$
via the Weierstrass representation.
\end{theorem}

{\sc Proof.}
Let $\hat{\Psi}$ and $\hat{\Phi}$ define the surface $S^{-1}(\Phi_0,\Psi_0)$ via the Weierstrass representation.
The formula (\ref{derivative}) implies the equality
$$
\hat{S} _z= -\frac{i}{2} \Gamma \hat{\Phi}_0^\top (1+\sigma_3)\hat{\Psi}_0=
\frac{i}{2} S^{-1}(\Phi_0,\Psi_0) \Gamma \Phi_0^\top (1+ \sigma_3) \Psi_0 S^{-1}(\Phi_0, \Psi_0),
$$
which is simplified up to the form
$$
\hat{\Phi}_0^\top (1+\sigma_3)\hat{\Psi}_0 =
-\Gamma^{-1} S^{-1} \Gamma \Phi_0^\top (1+ \sigma_3) \Psi_0 S^{-1}.
$$
It is easy to check the following identity
$$
-\Gamma^{-1} S^{-1}(\Phi_0, \Psi_0)\Gamma = \Gamma S^{-1}(\Phi_0, \Psi_0) \Gamma = (S^{-1}(\Psi_0, \Phi_0))^\top,
$$
which together with the preceding equality imply
\begin{equation}
\label{z}
D^\top (1+\sigma_3) C = (1+\sigma_3) = \left(\begin{array}{cc} 2 & 0 \\ 0 & 0 \end{array}\right)
\end{equation}
for
$$
C = \Psi_0 S^{-1}(\Phi_0, \Psi_0) \hat{\Psi}_0^{-1}, \ \ \
D = \Phi_0 S^{-1}(\Psi_0, \Phi_0) \hat{\Phi}_0^{-1}.
$$
Analogously, by considering $\hat{S}_{\bar{z}}$ and $S_{\bar{z}}$, we conclude that
$$
D^\top (\sigma_3-1)C = (\sigma_3-1) =  \left(\begin{array}{cc} 0 & 0 \\ 0 & -2 \end{array}\right).
$$
It follows from the last equality and from (\ref{z}) that the matrices $C$ and $D$ are diagonal and $C = D^{-1}$.
Since $\Psi_0, S^{-1}, \hat{\Psi}_0, \Phi_0,  \hat{\Phi}_0 \in \Q$, we have $C, D \in \Q$, therefore,
$$
C = D^{-1} = \left(\begin{array}{cc} e^{h} & 0 \\ 0 & e^{\bar{h}} \end{array}\right)
$$
and we infer that
$$
\hat{\Psi}_0 = \left(\begin{array}{cc} e^{h} & 0 \\ 0 & e^{\bar{h}} \end{array}\right)  \Psi_0 S^{-1}(\Phi_0, \Psi_0), \ \ \
\hat{\Phi}_0 = \left(\begin{array}{cc} e^{- h} & 0 \\ 0 & e^{- \bar{h}} \end{array}\right)  \Phi_0 S^{-1}(\Psi_0, \Phi_0),
$$
i.e. the spinors $(\hat{\Psi}_0,\hat{\Phi}_0)$ are obtained from $(\Psi_0 S^{-1}(\Phi_0, \Psi_0), \Phi_0 S^{-1}(\Phi_0, \Psi_0))$
by the gauge transformation (\ref{gauge}) and define the same surface.
Theorem 2 is proved.

For the completeness of exposition let us compute the function $W= \widetilde{U}-U$ in terms of the
Weierstrass representation.

\begin{proposition}
$$
W = \frac{\langle r, \p \rangle}{|r|^2} = \frac{e^\alpha}{|r|^2}\langle r, \n_1+i\n_2 \rangle,
$$
where $r:U \to \R^4$ is a surface in $\R^4$,
the vector $\p$ has the form (\ref{normal}), $e^{2\alpha}$ is the conformal factor of the metric and $(\n_1,\n_2)$ is a basis
of the normal bundle.
\end{proposition}

{\sc Proof.}
Let us compute the function $K_{22}=-iW$ given by (\ref{kmatrix}). We have
$$
K = \Psi_0 S^{-1}(\Phi_0, \Psi_0) \Gamma \Phi_0^\top \Gamma^{-1} =
$$
$$
\Psi_0 S^{-1} \left(\begin{array}{cc} 0 & 1 \\ -1 & 0 \end{array}\right)
\left(\begin{array}{cc} \varphi_1 & \varphi_2 \\ -\bar{\varphi}_2 & \bar{\varphi}_1 \end{array}\right)
\left(\begin{array}{cc} 0 & -1 \\ 1 & 0 \end{array}\right) =
$$
$$
=
\frac{1}{|r|^2}
\left(\begin{array}{cc} \psi_1 & -\bar{\psi}_2 \\ \psi_2 & \bar{\psi}_1 \end{array}\right)
\left(\begin{array}{cc} -ix^3 + x^4 & x^1 + ix^2 \\ -x^1 +ix^2 & ix^3 + x^4 \end{array}\right)
\left(\begin{array}{cc} \bar{\varphi}_1 & \bar{\varphi}_2 \\ -\varphi_2 & \varphi_1 \end{array}\right),
$$
where $|r|^2 = \sum_{k=1}^3 (x^k)^2$,
and conclude that, by (\ref{normal}),
$$
K_{22} = \frac{1}{|r|^2} (x^1
 (\psi_2\varphi_1 -\bar{\psi}_1\bar{\varphi}_2)+ ix^2 (\psi_2\varphi_1 +\bar{\psi}_1\bar{\varphi}_1)+ ix^3
(\bar{\psi}_1\varphi_1 - \psi_2 \bar{\varphi}_2) +
$$
$$
+ x^4 (\bar{\psi}_1 \varphi_1 + \psi_2 \bar{\varphi}_2) ) = - \frac{i}{|r|^2}  \langle r, \p\rangle.
$$
Proposition is proved.

\section{An integrable example of ``conformal'' transfor\-ma\-tions of the spectral curve and of the Floquet functions}

Let the potential $U$ is double-periodic:
$$
U(z+\lambda)=U(z), \ \ \ \lambda \in \Lambda \approx {\mathbb Z}^2 \subset \C.
$$
A solution $\psi$ of  (\ref{diraceq}) is called the Floquet function  (on the zero energy level) of
$\D$, if there exist constants $\mu_1$ and $\mu_2$ (the Floquet multipliers) such that
$$
\psi(z+\lambda_k) = \mu_k \psi(z), \ \ \ k=1,2,
$$
where $\lambda_1$ and $\lambda_2$ generate the period lattice $\Lambda$. The Floquet functions are parameterized by the spectral curve $\Gamma$ of $\D$ \cite{T98} (see, also, \cite{T-RS}), which was first introduced in
\cite{DKN} for the two-dimensional Schr\"odinger operator.

In \cite{GT}  it was proved that the actions of conformal transformations of $\R^4$ on tori preserve
the Floquet multipliers of the Dirac operators coming into their Weierstrass representations.
The proof consists in the following:

1) the identity map and the inversion are connected by a smooth curve $\gamma(t)$ in the space of conformal transformations;

2) to the torus $\Sigma \subset \R^2$ with a fixed conformal parameter $z$ was applied the conformal transformation $\gamma(t)$;

3) on the constructed torus $\Sigma_t = \gamma(t)\cdot \Sigma$ the parameter $z$ is also conformal and its Weierstrass representation
has the potential $U(z,\bar{z},t)$;

4) the derivatives in $t$ of the Floquet functions are computed and, therewith, it is proved that
the derivatives of the multipliers vanish.

Moreover in \cite{GT} it was shown that the evolution in  $t$ of the Floquet functions has the form
of a nonlinear equation of the Melnikov type.
In \cite{GT} it was pointed out that under such a deformation the spectral curve may become singular
due to creation of double points. We demonstrate that below by using explicit analytical formulas.

For tori in $\R^3$ the preservation of the multipliers, conjectured by us, was proved in \cite{GS}.
The question on the preservation of the spectral curve was not discussed in \cite{GS}.

Let us present an explicit example of such a deformation of a potential $U(z,\bar{z},t)$ and of the corresponding Floquet functions.

This examples is related to the Clifford torus $\Sigma$, which is defined by the equations
$$
(x^1)^2 + (x^2)^2 = \frac{1}{2}, \  \  ( x^3)^2 + (x^4)^2 = \frac{1}{2}
$$
and is parameterized as follows
$x^1=\frac{1}{\sqrt{2}} \cos x, x^2 = \frac{1}{\sqrt{2}} \sin x$, $x^3 = -\frac{1}{\sqrt{2}} \cos y$, $x^4 =  -\frac{1}{\sqrt{2}} \sin y$.
Its Weierstrass representation is given by the potential
\begin{equation}
\label{cliffordp}
U_{\rm clifford} =  -\frac{ i}{\sqrt{8}}
\end{equation}
and by the spinors
$$
\psi_0 = \frac{e^{-\frac{i(x+y)}{2}}}{\sqrt{2}}  \left(\begin{array}{c} e^{i\frac{3\pi}{8}} \\ e^{-i\frac{3\pi}{8}} \end{array}\right), \ \ \
\varphi_0 = \frac{e^{\frac{i(y-x)}{2}}}{2} \left(\begin{array}{c} -e^{-i\frac{3\pi}{8}} \\ e^{i\frac{3\pi}{8}} \end{array}\right)
$$
(in \cite{T-RS} we used the potential $U=\frac{1+i}{4}$ which is related to (\ref{cliffordp}) by a gauge transformation (\ref{gaugep})).

The basis of the Floquet functions of the Dirac operator (\ref{dirac}) with a constant potential $U$ may be taken in the form
$$
\psi(z, \bar{z}, \lambda) =\exp \left( \lambda z - \frac{|U|^2}{\lambda}\bar{z}\right) \left(\begin{array}{c} 1 \\ -\frac{U}{\lambda} \end{array}\right),
$$
where $\lambda \in \C \setminus \{0\}$ and the (compactified) spectral curve is the Riemann sphere: $\Gamma = \C \cup \{\infty\}$.
For the potential (\ref{cliffordp}) of the Clifford this basis takes the form
\begin{equation}
 \label{floquet}
\psi_{\rm Clifford}(z, \bar{z}, \lambda) = \exp \left( \lambda z -
\frac{1}{8\lambda}\bar{z}\right) \left(\begin{array}{c} 1 \\ \frac{i}{\sqrt{8}\lambda} \end{array}\right).
 \end{equation}

Let us consider the family of surfaces $\Sigma_t$ obtained from the Clifford tours by translations by $t$ along the $Ox^4$ where $t\in \R$:
$$
(x^1,x^2,x^3,x^4) \to (x^1,x^2,x^3,x^4+t),
$$
and apply to each torus form this family the mapping (\ref{stransform}), i.e. a composition of the inversion with the center at the origin and
the reflection (\ref{reflection}).
The obtained tori we denote by $\tilde{\Sigma}_t$. The potentials $U(z,\bar{z},t)$ of their Weierstrass representations are explicitly computed
by using Theorem 1 and are as follows:
$$
U(z,\bar{z},t)  =-\frac{i}{\sqrt{8}} + \frac{\sqrt{2}i - (1 + i) t \sin y }{2(t^2 - \sqrt{2}t \sin y + 1)},
$$

The tori $\tilde{\Sigma}_t$ are defined by the spinors
$$
\widetilde{\psi}_0 = \frac{e^{i\frac{3\pi}{8}}}{\sqrt{2}(t^2 -\sqrt{2}t \sin y + 1)}\left(\begin{array}{c} e^{i\frac{\pi}{4}} \exp \left(\frac{i(y-x)}{2}\right) + t  \exp \left(-\frac{i(x+y)}{2}\right) \\ \exp \left(\frac{i(y-x)}{2}\right)  - t e^{i\frac{\pi}{4}}\exp \left(-\frac{i(x+y)}{2}\right) \end{array}\right)
$$
$$
\widetilde{\varphi}_0 = \frac{e^{i\frac{3\pi}{8}}}{2 (t^2 -\sqrt{2}t \sin y + 1)}\left(\begin{array}{c} -\exp \left(-\frac{i(x+y)}{2}\right) - t e^{i\frac{\pi}{4}}\exp \left(\frac{i(y-x)}{2}\right) \\ e^{i\frac{\pi}{4}}\exp \left(-\frac{i(x+y)}{2}\right) - t  \exp \left(\frac{i(y-x)}{2}\right) \end{array}\right)
$$

We derive from these formulas that

\begin{enumerate}
\item
for $t=0$ the Clifford torus is mapped into itself and the potential is mapped into a gauge equivalent potential:
$$
U_{\rm Clifford} = - \frac{i}{\sqrt{8}} \to U(z,\bar{z},0) = \frac{i}{\sqrt{8}};
$$

\item
only for $t=\pm 1$ the spinors $\tilde{\psi}$ and $\tilde{\varphi}$ are proportional:
$$
\tilde{\psi} = \mp \sqrt{2} \tilde{\varphi},
$$
or, which is equivalent, the surface lies in the three-dimensional hyper\-pla\-ne.
Indeed, only in these cases the torus
$\Sigma_t$ passes through the origin and by the inversion is mapped into the hyperplane
$x^4 = {\rm const}$.
In this hyperplane the surface  $\tilde{\Sigma}_t$ is the Clifford torus (in $\R^3$) on which the Willmore functional attains its minimum among all tori in
$\R^3$ \cite{MN}. The potentials of these tori are equal to
$$
U(z,\bar{z},\pm 1) = \mp \frac{\sin y}{2\sqrt{2}(\sqrt{2} \mp \sin y)};
$$

\item
by Theorem 1, the Floquet functions of the operator with the potential $U(z,\bar{z},t)$ are obtained from the functions (\ref{floquet}) by the Moutard transformation
and take the form
$$
\widetilde{\psi} =
\exp \left( \lambda z - \frac{1}{8\lambda}\bar{z} \right)
\left(\begin{array}{c} 1 - \frac{2 i + 2 t \cdot ( 2\sqrt{2} i \lambda \cos y + e^{- i \frac{\pi}{4}} \sin y)}{(8 \lambda ^2+i)(t^2 - \sqrt{2}t \sin y + 1)}
\\  i\left( \frac{1}{\sqrt{8}\lambda} + \frac{-4\sqrt{2} \lambda + 2 t \cdot (\cos y + 2\sqrt{2} e^{-i \frac{\pi}{4}} \lambda \sin y) }{(8 \lambda ^2+i)(t^2 - \sqrt{2}t \sin y + 1)} \right)
\end{array}\right).
$$
By multiplying that by $\lambda^8+i$ to get rid of the appearing poles in $\lambda$ and by saving, for brevity, the notations, we
derive for
$$
u = \frac{1+i}{4}, \ \ \
t=1
$$
that
$$
\widetilde{\psi}(u) =
 \left(\begin{array}{c} \frac{i \sqrt{2} e^{i \frac{x+y}{2}} + (1 - i )e^{i \frac{x-y}{2}}}{  \sqrt{2} -  \sin y } \\ \frac{ (i - 1)e^{i \frac{x+y}{2}} + i \sqrt{2}e^{i \frac{x-y}{2}}}{\sqrt{2} -  \sin y} \end{array}\right),
 \ \
\widetilde{\psi}(-\bar{u})=
 \left(\begin{array}{c} \frac{(1 + i) e^{i \frac{x+y}{2}} - i \sqrt{2} e^{i \frac{x-y}{2}}}{  \sqrt{2} -  \sin y } \\ \frac{i \sqrt{2}e^{i \frac{x+y}{2}} + (1 + i)e^{i \frac{x-y}{2}}}{\sqrt{2} -  \sin y} \end{array}\right),
$$
$$
\widetilde{\psi}(-u) =
 \left(\begin{array}{c} \frac{i \sqrt{2} e^{-i \frac{x+y}{2}} + ( i - 1 )e^{i \frac{y - x}{2}}}{  \sqrt{2} -  \sin y } \\ \frac{ (1 - i)e^{-i \frac{x+y}{2}} + i \sqrt{2}e^{i \frac{y - x}{2}}}{\sqrt{2} -  \sin y} \end{array}\right),
\ \
\widetilde{\psi}(\bar{u}) =
 \left(\begin{array}{c} \frac{-(1 + i) e^{-i \frac{x+y}{2}} - i \sqrt{2} e^{i \frac{y-x}{2}}}{  \sqrt{2} -  \sin y } \\ \frac{i \sqrt{2}e^{-i \frac{x+y}{2}} - (1 + i)e^{i \frac{y-x}{2}}}{\sqrt{2} -  \sin y} \end{array}\right),
$$
which implies the following equalities
$$
\widetilde{\psi}(u) =  \frac{1+i}{\sqrt{2}} \, \widetilde{\psi}(-\bar{u}), \ \
\widetilde{\psi}(-u) = -\frac{1+i}{\sqrt{2}} \, \widetilde{\psi}(\bar{u}).
$$
Hence, for $t=1$ the Floquet functions are uniquely parameterized
by points of the singular curve $\C \setminus \{0\}/\{u \sim -\bar{u}, -u \sim \bar{u}\}$, and the spectral curve $\Gamma_1$, compactified by a pair of
``infinities'' $\lambda=0$ and $\lambda=\infty$, is a rational curve with a pair of double points.
These finite gap integration data for the Clifford torus were obtained in \cite{TC}.

From the explicit formulas for the Floquet functions it is easy to notice that for small $t$ the spectral curve of the operator with the potential
$U(z,\bar{z},t)$ is preserved and stays smooth, and for $t=1$ on it appear a pair of double points.
Therewith the Floquet multipliers are preserved.
\end{enumerate}


\begin{thebibliography}{MMM}

\bibitem{T151}
Taimanov, I.A.:
The Moutard transformation of two-dimensional Dirac operators and the Mobius geometry.
Math. Notes {\bf 97}:1 (2015), 124--135.

\bibitem{C}
Delong Yu, Q.P. Liu, and Shikun Wang:
Darboux transformation for the modified Veselov--Novikov equation.
J. of Physics A {\bf 35} (2001), 3779--3785.

\bibitem{TDS}
Taimanov, I.A.:
Surfaces in the four-space and the Davey-Stewartson equations.
J. Geom. Phys. {\bf 56}:8 (2006), 1235--1256.

\bibitem{T-RS}
Taimanov, I.A.
Two-dimensional Dirac operator and surface theory.
Russian Math. Surveys {\bf 61}:1 (2006), 79--159.

\bibitem{K2}
Konopelchenko, B.G.:
Weierstrass representations for surfaces in 4D spaces and their integrable deformations via DS hierarchy.
Ann. Global Anal. Geom. {\bf 16}:1 (2000), 61--74.

\bibitem{T-MNV}
Taimanov, I.A.:
Modified Novikov--Veselov equation and
differential geometry of surfaces.
Amer. Math. Soc. Transl., Ser. 2,
V. 179, 1997, pp. 133--151.

\bibitem{T98}
Taimanov, I.A.:
The Weierstrass representation of closed surfaces in $\R^3$.
Functional Anal. Appl. 32:4 (1998), 258-267.

\bibitem{DKN}
Dubrovin, B.A., Krichever, I.M., and Novikov, S.P.:
The Schr\''odinger equation in a periodic field and Riemann surfaces. Sov. Math. Dokl. {\bf 17} (1976), 947--951.

\bibitem{GT}
Grinevich, P.G., and Taimanov, I.A.:
Infinitesimal Darboux transformations of the spectral curves of tori in the four-space.
Int. Math. Res. Not. 2007, no. 2, Art. ID rnm005.

\bibitem{GS}
Grinevich, P.G., and Schmidt, M.U.:
Conformal invariant functionals of tori into $\R^3$.
J. Geom. Phys. {\bf 26} (1998), 51--78.

\bibitem{MN}
Marques, F.C., and Neves, A.:
Min-Max theory and the Willmore conjecture.
Ann. Math. {\bf 179} (2014), 683--782.

\bibitem{TC}
Taimanov, I.A.:
Finite gap theory of the Clifford torus.
International Mathematics Research Notices (2005), 103--120.



\end{thebibliography}
\end{document}